\documentstyle[twocolumn,epsfig]{article}
\begin{document}

\title{\large{Nonet Classification of Scalar/Isoscalar Resonances
in the Mass Region below\\ 1900 MeV:
Observation of the Lightest Scalar Glueball}}
\author{\small{A.V. Anisovich$^1$, \underline{V.V. Anisovich}$^1$,
Yu.D. Prokoshkin$^2$ and A.V. Sarantsev$^1$}\\
\small{(1) Petersburg Nuclear Physics Institute, Gatchina,
St.Petersburg 188350, Russia}\\
\small{(2) Institut for High Energy Physics, Protvino, Serpukhov,
142284, Russia}}
\date{}
\maketitle

Here we summarize the results of the investigation of the $IJ^{PC}=00^{++}$
wave [1-3]: this investigation is devoted to the search for the
lightest scalar glueball. In Refs. \cite{km1550,km1900},
the $q\bar q$-nonet
classification of the scalar/isoscalar states is performed in the mass
region below 1900~MeV basing on the following data: GAMS data for
$\pi^-p\to\pi^0\pi^0n$ \cite{gams1}, $\eta\eta n$ \cite{gams2},
$\eta\eta'n$ \cite{gams3};
CERN-M\"unich data for $\pi^-p\to\pi^+\pi^-n$ \cite{cern};
Crystal Barrel data
for $p\bar p\to\pi^0\pi^0\pi^0$, $\pi^0\pi^0\eta$,
$\pi^0\eta\eta$ \cite{cbc1,cbc2}; BNL data for $\pi\pi\to K\overline K$
\cite{bnl}. The
nonet classification of the states and search for  extra states look
like the only way for the identification of  lightest glueballs. The
$00^{++}$-wave is analyzed in Refs. \cite{km1550,km1900}
in terms of the $K$-matrix
elements which are presented in the form:
\begin{equation}
K_{ij}\ =\ \sum_\alpha\frac{g^{(\alpha)}_ig^{(\alpha)}_j}{
m^2_\alpha-s}\ +\ \mbox{ smooth terms },
\end{equation}
where the pole terms ($s$ is invariant energy squared) describe the
input meson states "before" the decay and their mixture due to the
transition into $\pi\pi$, $K\overline K$, $\eta\eta$, $\eta\eta'$ and
$\pi\pi\pi\pi$; the $K$-matrix poles are referred as poles of "bare
states". The couplings $g^{(\alpha)}$ are related to the input-state
couplings of
 the dispersion relation $N/D$-amplitude: the
$K$-matrix couplings obey the same relations as  couplings of
input poles of the $N/D$-amplitude \cite{aas}, thus allowing to analyse
 the $q\bar q$/glueball content of bare states. Coupling
constants are determined by leading terms in the $1/N_c$-expansion for
the process of Fig. 1: gluons  produce new $q\bar q$-pairs with the
flavour symmetry violation, $u\bar u:d\bar d:s\bar s=1:1:\lambda$, with
strange-quark production probability suppression factor $\lambda\simeq0.5$
\cite{km1900}.
The $q\bar q$-state with non-strange/strange quark content $q\bar
q=n\bar n\cos\phi+s\bar s\sin\phi$, where $n\bar n=(u\bar u+d\bar
d)/\sqrt2$,
\epsfig{file=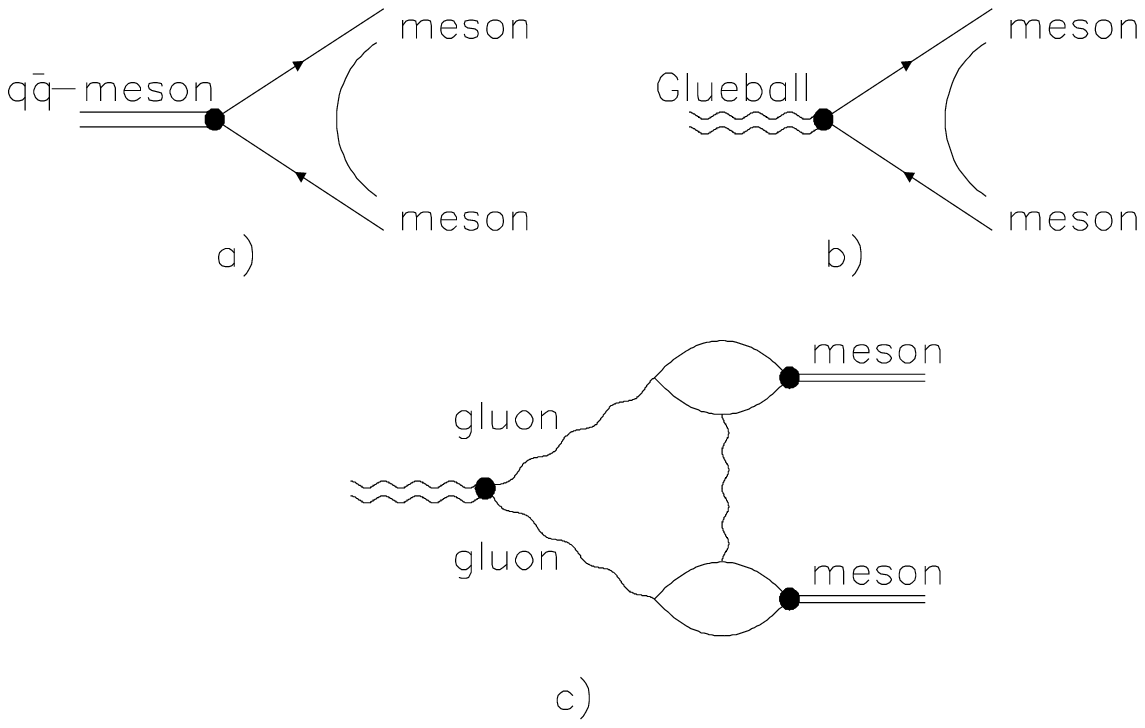,width=8cm,rheight=5cm}
\begin{description}
\item[] Figure 1. Diagrams for the decay of a
$q\bar q$-meson (a) and a glueball (b,c) into two $q\bar q$-meson
states.
\end{description}
decays into two pseudoscalar mesons (Fig.1a) with the
following normalized production probabilities \cite{km1550,km1900}:
\begin{eqnarray}
 &&\gamma^2_{\pi\pi}:\gamma^2_{K\bar
K}:\gamma^2_{\eta\eta}:\gamma^2_{\eta\eta'}\ =\nonumber\\
&=&\frac32\cos^2\phi:\frac12\left(\sqrt2\sin\phi+\sqrt\lambda\cos\phi
\right)^2\ : \nonumber \\
&:&\frac12\left(\cos^2\theta\cos\phi+\sqrt{2\lambda}\sin^2\theta
\sin\phi\right)^2 \nonumber \\
&:&\sin^2\theta\cos^2\theta\left(\cos\phi-
\sqrt{2\lambda}\sin\phi\right)^2\ .
\end{eqnarray}
Here $\gamma^2_i=\Gamma_i/\Omega_i$, where $\Gamma_i$ is  partial
width and $\Omega_i$ is  corresponding phase space factor. The
angle $\theta$ determines the quark content of $\eta$ and $\eta'$.
The glueball decay into two pseudoscalar mesons, being a two-meson
 production process (Fig.1b), has the same
couplings as given in
Eq.(2) but with fixed angle $\phi\to\phi_{\rm glueball}$:
\begin{equation}
\tan\phi_{\rm glueball}\ =\ \sqrt{\frac\lambda 2}\ .
\end{equation}
\epsfig{file=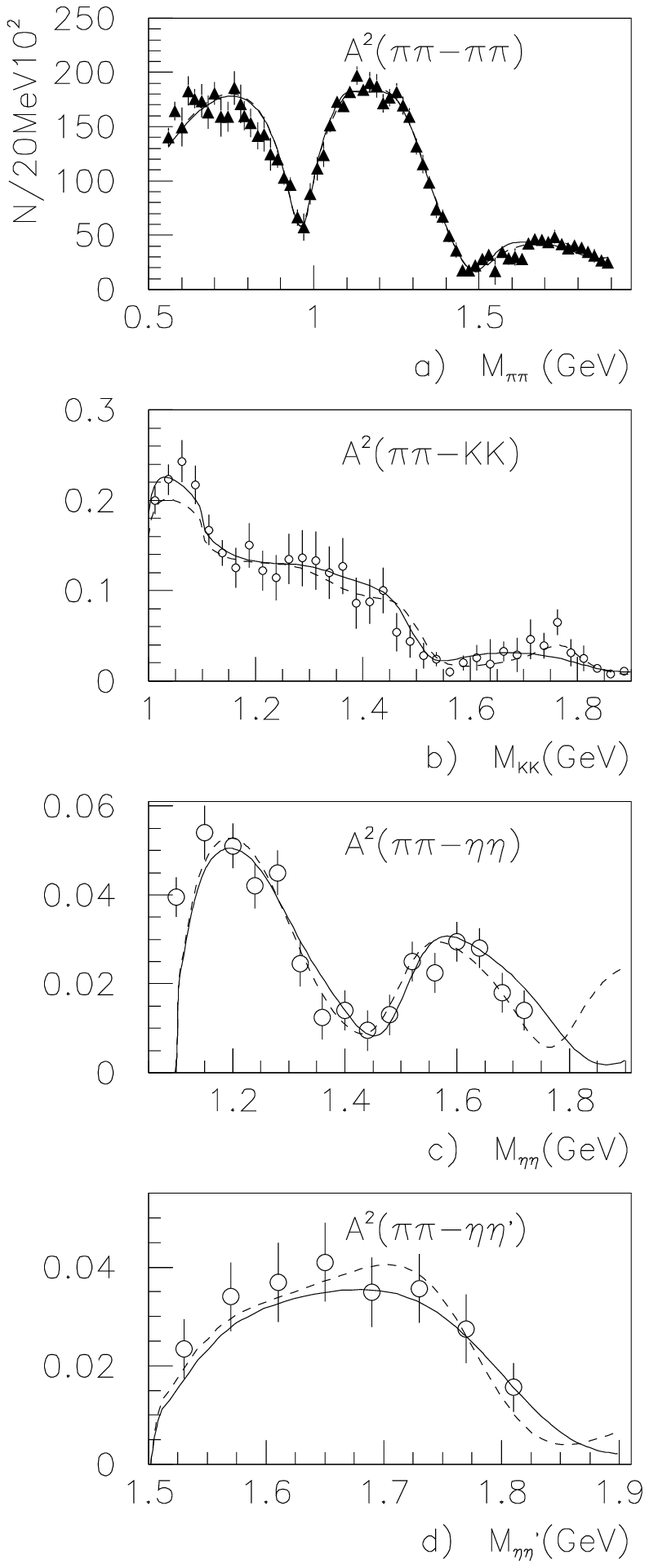,width=7cm,rheight=17cm,bbllx=0cm,bblly=3.5cm,
bburx=8cm,bbury=20.5cm}
\begin{description}
\item [] Figure 2. Description of the $S$-wave amplitude
squared in the reactions:
 $\pi\pi\to \pi\pi$ (data are taken from ref. \cite{gams1}) (a),
 $\pi\pi\to K\bar K$ \cite{bnl} (b),  $\pi\pi\to \eta\eta$
\cite{gams2} (c) and  $\pi\pi\to \eta\eta'$ \cite{gams3} (d).
Solid curves correspond to solution {\bf II} and dashed curves
to  solution {\bf I}.
\end{description}
With Eqs. (2) and (3) for the couplings, a simultaneous fit of the
two-meson spectra of Refs. [4--10] was performed \cite{km1550,km1900}
in the framework
of the $K$-matrix representation of the amplitudes. Two solutions
describe well the data set:

{\bf Solution I.} Two bare states $f^{bare}_0(720\pm100)$ and
$f^{bare}_0(1260\pm30)$ are members of the $1^3P_0$ $q\bar q$-nonet, with
$f^{bare}_0(720)$ being $s\bar s$-rich state: $\phi(720)=-69^o\pm12^o$.
 The bare states $f^{bare}_0(1600\pm50)$ and $f^{bare}_0
(1810\pm30)$ are  members of the $2^3P_0$-nonet; $f^{bare}_0
(1600)$ is dominantly $n\bar n$-state: $\phi(1600)=-6^o\pm15^o$. The
state $f^{bare}_0(1235\pm50)$ is superfluous for $q\bar
q$-classification, being a candidate for the lightest glueball: its
couplings to the two-\-meson decay obey Eq. (3).

{\bf Solution II.} Basic nonet members are the same as in
solution I. The members of the $2^3P_0$-\-nonet are the following:
$f^{bare}_0(1235)$ and $f^{bare}_0(1810)$; both these states
have significant $s\bar s$-component: $\phi(1235)=42^o\pm10^o$ and
$\phi(1810)=-53^o\pm10^o$. The state $f^{bare}_0(1560\pm30)$ is
supersluous for the $q\bar q$-\-classification, being a good candidate for
the lightest glueball.

The description of  meson spectra taken from Refs. [4-10] using the solutions I
and II is shown in Figs. 2-5.
In the region 1000--1900 MeV there are five scalar resonances
which are related to the five poles in the complex-mass
plane:
\begin{equation}
\begin{array}{cl}
\mbox{Resonance } & \mbox{ Position of  poles on the }\\
& \mbox{ complex-$M$ plane, in MeV units }\\ &\\

f_0(980) & 1015\pm15-i(43\pm8)\\
f_0(1300) & 1300\pm20-i(120\pm20)\\
f_0(1500) & 1499\pm8-i(65\pm10) \\
f_0(1750) & 1780\pm30-i(125\pm70)\\
f_0(1200-1600) & 1530^{+90}_{-250}-i(560\pm140)\ .
\end{array}
\end{equation}
Comparison of the positions of  bare states
and those of real resonances
shows that the observable states
are strong mixtures of  bare states
and two/four-meson states (into which these resonances decay).

For further analysis of the mixing,
we have rewritten the $00^{++}$-wave amplitude in terms of the
$q\bar q$-states and input bare states
 \cite{aas}. Although this transition from
the $K$-matrix to the amplitude, which is analytic in the
whole $s$-plane,
faces ambiguities
 related to the left-\-hand side singularities of
vertices $bare\;state \to q\bar q$, $g_a(s)$,
 it allows us to restore the mass of the pure glueball,
that is an object of lattice calculations. In addition, it
clarifies the situation with the mixture of overlapping resonances.
The point
is that in the case of two overlapping and mixing resonances, one resonance
accumulates the widths of both initial states
($\Gamma_{broad}\simeq
\Gamma_1+\Gamma_2)$, while the other resonance be-
\epsfig{file=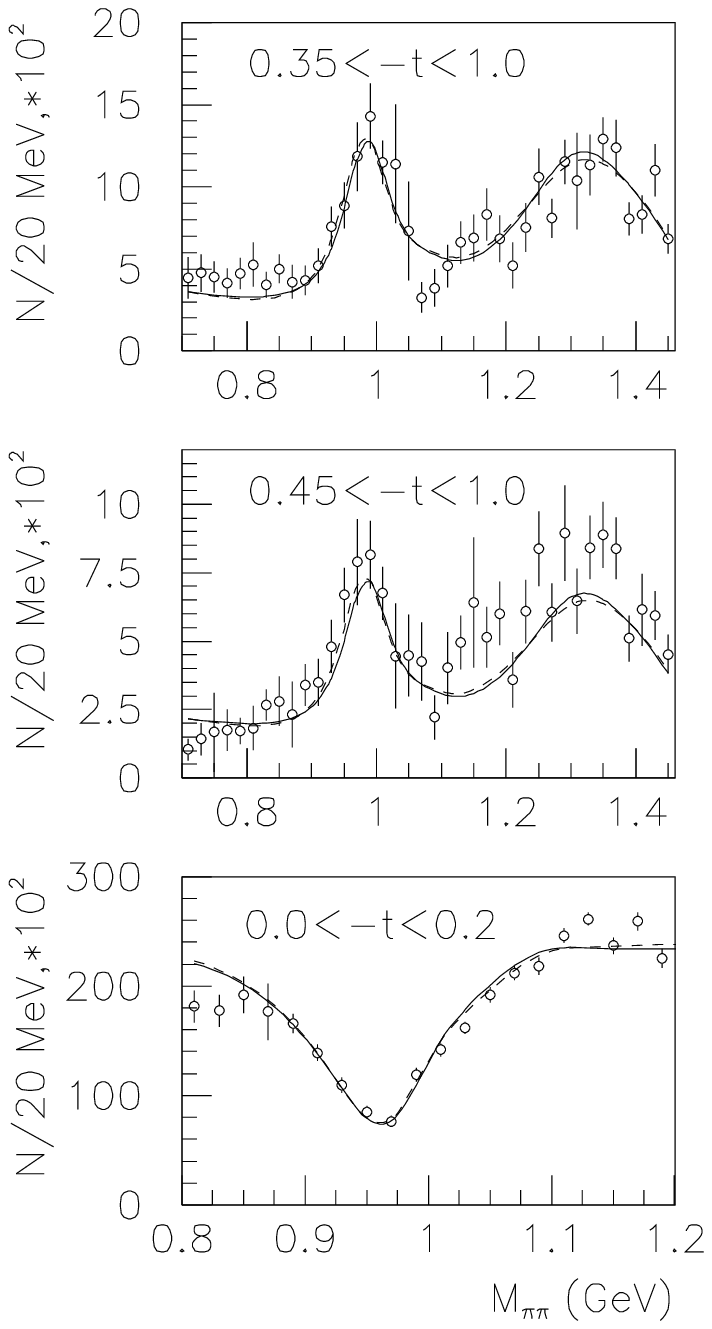,width=8cm,rheight=14.5cm,
bbllx=0cm,bblly=0.5cm,bburx=8cm,bbury=15cm}
\begin{description}
\item[] Figure 3. Event
numbers {\it versus}  invariant mass of the $\pi\pi$-system for different
$t$-intervals in the $\pi^-p\rightarrow \pi^0\pi^0n$ reaction
\cite{gams1}. Solid curves correspond to solution {\bf{II}}
and  dashed curves to solution {\bf{I}}.
\end{description}
comes narrow
$(\Gamma_{narrow}\to0)$. In the case of three overlapping resonances,
two final
states have small widths ($\Gamma_{\rm narrow-1}\to0$, $\Gamma_{\rm
narrow-2}\to0)$,
\epsfig{file=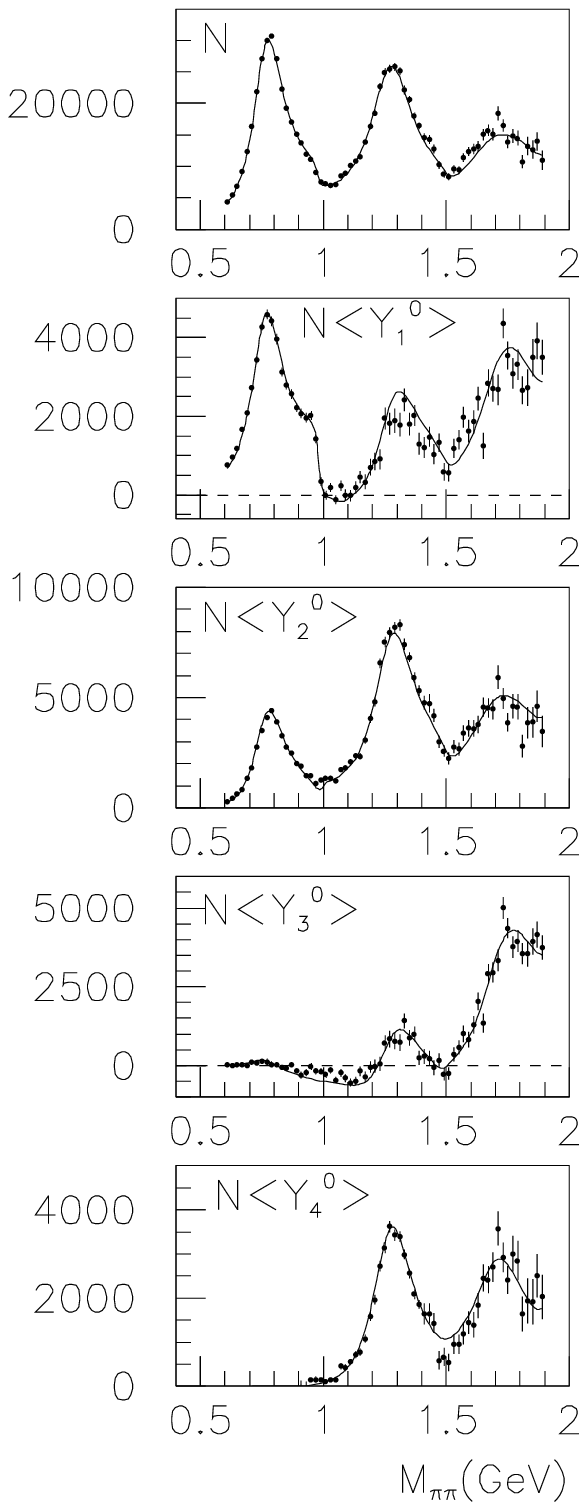,width=8.5cm,
bbllx=0cm,bblly=1.0cm,bburx=8cm,bbury=16.8cm}
\begin{description}
\item[] Figure 4.  Fit of the $\pi\pi$ angular-moment
distributions   in the final state of the reaction
$\pi^-p \to n\pi^+\pi^-$ at 17.2 GeV/c \cite{cern}. The curve
corresponds to solution {\bf{II}}.
\end{description}
while the third resonance accumulates the widths of all initial
states $(\Gamma_{\rm broad}\simeq \Gamma_1+\Gamma_2+\Gamma_2)$.

The glueball propagator, which takes into account the transitions
of Fig.6 type, is determined
\epsfig{file=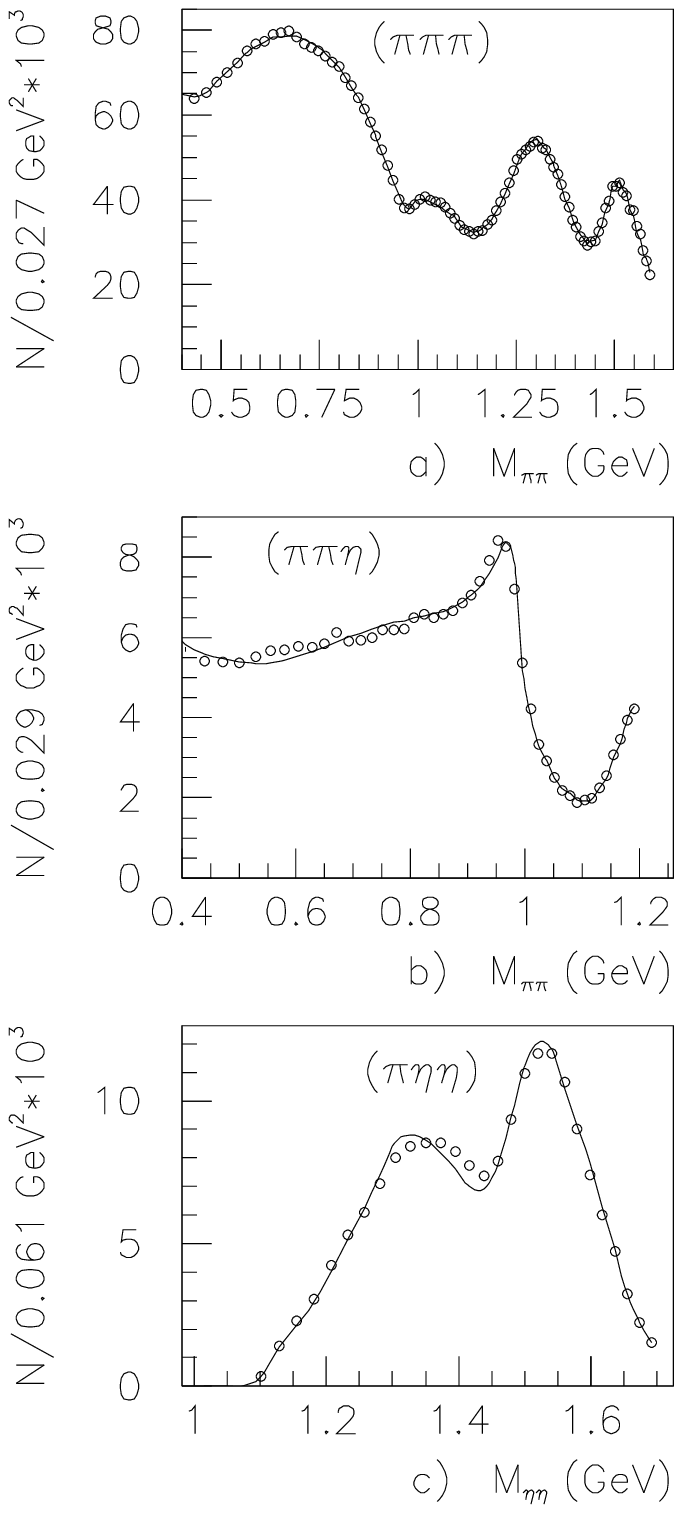,width=8cm,
bbllx=0cm,bblly=0.5cm,bburx=8cm,bbury=17cm}
\begin{description}
\item[] Figure 5. The $\pi^0\pi^0$ spectra: (a) in the $p\bar p\to
\pi^0\pi^0\pi^0$ reaction,  (b) in the $p\bar p\to
\eta\pi^0\pi^0$ reaction; (c) $\eta\eta$ spectra in the $p\bar p\to
\pi^0\eta\eta$ reaction.
Curves correspond to solution {\bf{II}}.
\end{description}
by the $q\bar q$ loop diagrams with the vertices $g_a(s)$:
\begin{equation}
B_{ab}(s)=\int\limits_0^1 \frac{dx}{x}\int\frac{d^2k_\perp}{(2\pi)^3}
\;\frac{g_a(s')g_b(s')}{s'-s}\;2(s'-4m^2)\;,
\end{equation}
where $s'=\frac{m^2+k_\perp^2}{x(1-x)}$ and $m$ is the quark mass.
Two types of quark loop diagrams are taken into account:
with non-strange $(n\bar n)$ and strange $(s\bar s)$ quarks, their relative
weights are given by mixing angles   $\phi$.
Complex masses of the physical states are determined as zeros of the
determinant:
\begin{equation}
det\left |(m_a^2-s)\delta_{ab}-B_{ab}(s)\right |=0\;,
\end{equation}
where $m_a$ are the masses of the input meson states and $\delta_{ab}$
is unit matrix.

\epsfig{file=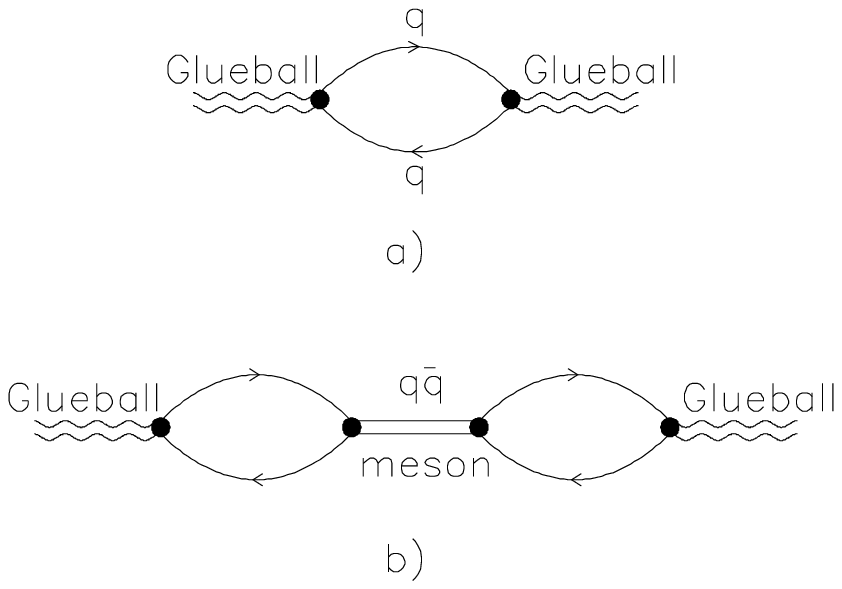,width=5.1cm,
bbllx=0cm,bblly=4.2cm,bburx=8cm,bbury=10.5cm}
\begin{description}
\item[] Figure 6. Diagrams for mixing glueball/$q\bar q$ mesons.
\end{description}
\vskip 0.2cm

Within a simple ansatz about the power decrease of vertices $g_a(s)$
at  large s, we refit the $00^{++}$ amplitude in the mass region
1200 -1600 MeV. Results for  solutions I and II are shown in Fig. 7.
To illustrate the dynamics of mixing, the following method is used: we
change $g_a\to \xi g_a$ and vary $\xi$ from 0 (no mixing)
 to 1 (the real state corresponding to the description of  data).
The pole movement with $\xi$ is shown in Fig. 7.
For both solutions the broad resonance $f_0(1530^{+90}_{-250})$ is
the descendant of the pure glueball. No wonder: the glueball
mixes with both $q\bar q$-states without suppression, while the mixture
of the $q\bar q$-states is suppressed, for they are members of
different $q\bar q$-nonets. Final position of  poles after the mixture of
initial (or bare) states supports the idea about the mixture of a glueball
with neighbouring $q\bar q$-mesons. The glueball, being a particle of
another origin than  $^3P_0~q\bar q$-states, sets foot in the
series of $q\bar q$ states and mixes with neightbouring mesons: an
existence of a broad state is an inevitable result of such an "invasion".

The broad resonance (the descendant of the pure glueball) has the
following $q\bar q$-meson/glueball content in solutions I and II
correspondingly:
$f_0(1530^{+90}_{-250})\to43\%(q\bar q)_1$ +
$28\%(q\bar q)_2$ + $29\%$ $G$;\\
$f_0(1530^{+90}_{-250})\to10\%(q\bar q)_1$ +
$42\%(q\bar q)_2$ + $48\%$ $G$.\\
The mass of the pure glueball in the solution II ($\xi=0$) is
equal to:
\begin{equation}
m_{pure~glueball}=1695 MeV
\end{equation}
This value agrees with  the results of  lattice
gluodynamics: $1600\pm 85\pm 100$ MeV \cite{close},
$1740\pm 70$ MeV \cite{ibm}.
Present analysis allows to conclude:
in the region 1300-1600~MeV we have the
scalar glueball. The pure glueball
state has mainly dispersed  over three real
resonances: $f_0(1300)$, $f_0(1500)$ and $f_0(1530^{+90}_{-250})$.
\epsfig{file=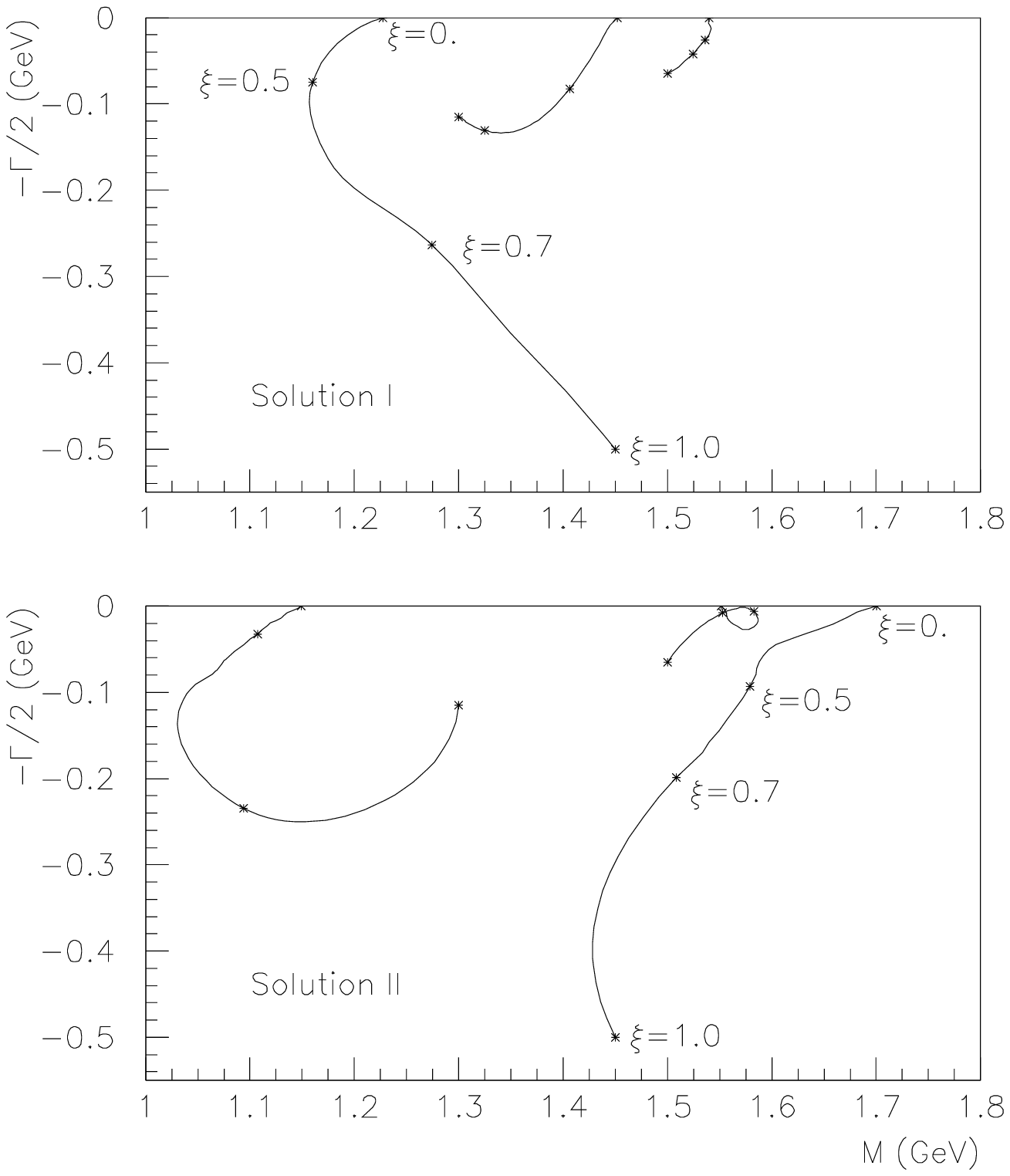,width=8.cm,rheight=9.5cm}
\begin{description}
\item[] Figure 7. Trajectories of poles in the complex
$M$-plane with increase of coupling constants: $g_a\to
\xi g_a$.
\end{description}

\end{document}